\documentclass[letterpaper,twocolumn,10pt]{article}
\usepackage{usenix2019_v3}

\usepackage{amsmath,amssymb}
\usepackage{array}
\usepackage{booktabs}
\usepackage{enumitem}
\usepackage{graphicx}
\usepackage{xspace}
\usepackage{pifont}
\usepackage{algorithm}
\usepackage{algorithmic}

\SetSymbolFont{symbols}{normal}{OMS}{cmsy}{m}{n}
\SetSymbolFont{symbols}{bold}{OMS}{cmsy}{b}{n}
\DeclareMathAlphabet{\mathcal}{OMS}{cmsy}{m}{n}

\graphicspath{{}}

\definecolor{ssText}{HTML}{111820}
\definecolor{ssMuted}{HTML}{5B6670}
\definecolor{ssBase}{HTML}{5AA2CF}
\definecolor{ssBaseFill}{HTML}{DCEBF5}
\definecolor{ssOurs}{HTML}{F2C14E}
\definecolor{ssOursDark}{HTML}{8A5A00}
\definecolor{ssOursFill}{HTML}{FFF2BF}
\definecolor{ssGreen}{HTML}{4F7868}
\definecolor{ssGreenFill}{HTML}{EDF3F0}
\definecolor{ssNeutral}{HTML}{EDF1F4}
\definecolor{ssRed}{HTML}{B5533C}

\newcommand{\sys}{\textsc{ShardFly}\xspace}

\newcommand{\cmark}{\textcolor{ssGreen}{\ding{51}}}
\newcommand{\xmark}{\textcolor{ssRed}{\ding{55}}}
\newcommand{\pmark}{\textcolor{ssOursDark}{$\triangle$}}
\newcommand{\nmark}{\textcolor{ssMuted}{--}}
\newcommand{\para}[1]{\noindent\textbf{#1}\hspace{0.4em}}

\title{\Large \bf When Shards Fly: Cross-Tensor In-Flight Conversion\\for Low-Latency, Low-Memory Weight Updates in Disaggregated RL}
\author{%
  Jiayu Xiao\textsuperscript{1,2}\quad Ye Wang\textsuperscript{1}\quad
  Zhi Zhang\textsuperscript{2}\quad Jiarui Guo\textsuperscript{2}\quad
  Kan Liu\textsuperscript{2}\quad Tao Lan\textsuperscript{2}\quad
  Minchen Yu\textsuperscript{1}\thanks{Corresponding author:
  \texttt{yuminchen@cuhk.edu.cn}.} \\[0.6em]
  {\normalsize\textrm{\textsuperscript{1}The Chinese University of Hong Kong,
   Shenzhen}} \\
  {\normalsize\textrm{\textsuperscript{2}Alibaba Group}}}
\date{}

\hypersetup{pdftitle={When Shards Fly: Cross-Tensor In-Flight Conversion for Low-Latency, Low-Memory Weight Updates in Disaggregated RL},
  pdfauthor={Jiayu Xiao, Ye Wang, Zhi Zhang, Jiarui Guo, Kan Liu, Tao Lan, Minchen Yu}}

\begin{document}

\maketitle

\begin{abstract}
Disaggregated LLM reinforcement learning (RL) repeatedly synchronizes
policy weights between training and rollout GPU pools.
Existing systems reduce communication through shard delivery and replica
forwarding, yet separate conversion passes and staging buffers still
add latency and GPU-memory overheads to synchronous updates.
We present \sys, which coordinates reusable transfer planning with
streaming GPU execution for synchronous, in-place weight updates.
\sys introduces two components. The Planner compiles native shard mappings, quantization dependencies,
and replica routes into cross-tensor schedules, reserving payload
buffers only for scheduled transfers. 
The Executor runs each rank's schedule in one persistent transfer kernel,
fusing memory-layout conversion, quantization, GPU-initiated
communication, and destination writes while forwarding transformed
bytes across replicas.
Plans and communication state are reused under stable configurations
and replaced between updates for coordinated rollout scaling.
On up to 32 H20 GPUs, \sys achieves complete BF16 weight-update
speedups of 1.26--27.92$\times$ across five models and four baselines:
veRL NCCL, Awex, veRL NIXL, and WeightBridge.
In end-to-end Qwen3-30B-A3B GRPO across four rollout configurations,
\sys improves response-token throughput by
5.49--6.71\% over veRL NCCL.

\end{abstract}

\section{Introduction}
\label{sec:intro}

Disaggregated LLM reinforcement learning (RL) runs policy optimization
and rollout on separate GPU pools, allowing each stage to use suitable
parallelism configurations and execution
engines~\cite{hu2025openrlhf,zhong2025streamrl}.
Repeated \emph{weight updates} propagate optimized policy parameters
from trainers to rollout engines.
Larger sparse mixture-of-experts (MoE) models increase full-weight
transfer volumes relative to per-token
computation~\cite{deepseek2026v4,qwen2026qwen35}, while GPU compute
throughput improves faster than inter-node RDMA
bandwidth~\cite{han2025vclos,jiang2026weightbridge}.
This makes weight synchronization an increasingly important overhead:
WeightBridge reports GPUs stalled on weight transfer for up to 27\%
of production execution time~\cite{jiang2026weightbridge}.

A common approach uses \emph{synchronous, in-place weight
updates}.
Trainers pause optimization, and rollout engines pause generation while
new weights overwrite existing parameter tensors~\cite{wu2025llamarl,jiang2026weightbridge}.
Both pools wait until every rollout engine has installed the complete
new version, so the slowest destination determines the stall.
Complete update latency, including communication, conversion, and
destination writes, therefore lies on the RL execution critical path.
This update mode also applies within asynchronous RL workflows.

Efficient weight updates should meet four coupled requirements: (1) reducing redundant communication, (2) sustaining throughput with a small communication-memory footprint, (3) integrating parameter conversion with transfer, and (4) adapting to rollout dynamics. 
Existing systems provide shard-specific delivery and replica
forwarding~\cite{inclusionai2025awex,kandadi2026ncclm2n},
overlap preparation with RDMA~\cite{licker2026fabriclib}, or rebuild
communication topology as the rollout pool
changes~\cite{sheng2025hybridflow}.
The studied paths do not demonstrate all four capabilities together. 
For example, existing communication optimizations can rely on large staging buffers,
separate conversion stages add intermediate writes and rereads,
and rollout changes require routing, buffering, and execution state
to be updated consistently.
These interactions call for coordinated planning and execution across
the complete update path.


We present \sys, which combines reusable transfer planning with streaming GPU
execution through a \emph{Planner} and an \emph{Executor}. \sys exploits the
synchronous update contract: source weights remain unchanged, and rollout
engines do not read destination parameters during an update. Independent
transfers can therefore be scheduled across tensor boundaries, subject to
conversion and communication dependencies. The Planner compiles these
requirements into reusable schedules, and the Executor streams updates into
rollout parameter storage.

The Planner coordinates shard delivery, replica reuse, and
communication-buffer allocation. It maps source--destination shard overlaps to
native parameter fragments and builds replica routes that share trainer
injections and exploit local NVLink connectivity. Cross-tensor schedules
distribute transfer work across channels, while buffer allocations follow the
schedule to reduce unused payload reservations and preserve scheduled
concurrency. Prepared plans and communication state are reused while
descriptors and placement are stable. Coordinated rollout changes trigger
replacement plans and communication state prepared between updates to adapt
the update path to the new destination pool.

The Executor runs each rank's schedule in one persistent transfer kernel. It
fuses native-fragment gathering, memory-layout conversion, and quantization
with GPU-initiated transfer, streaming weights and scales into rollout
parameter tensors. Relays forward the same transformed bytes, so replicas
reuse the prepared representation. Using GPU resources available during the
pause, the pipeline overlaps preparation, communication, forwarding, and
destination writes. This reduces per-operation host dispatch, CPU--GPU control
round trips, and intermediate memory traffic.

\begin{figure}[!t]
  \centering
  \includegraphics[width=\columnwidth,trim=0 1.75bp 0 0,clip]{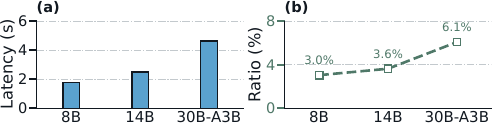}\par
  \caption{Recurrent weight-update time and its share of a complete
  trainer update cycle (veRL NCCL). Workload and layout:
  Appendix~\ref{sec:reproducibility}.}
  \label{fig:sync-model-size}
\end{figure}

We evaluate \sys across five models and multiple deployment configurations
on up to 32 H20 GPUs.
Across seven BF16 configurations and four baselines (veRL NCCL, Awex,
veRL NIXL, and WeightBridge), complete weight-update speedups span
1.26--27.92$\times$.
In a separate Qwen3-30B-A3B memory study, \sys completes a BF16 update
in 335\,ms with 0.91\,GiB/GPU of additional resident memory, a measured
footprint 60.9--82.8\% smaller than the selected baselines at their
fastest measured configurations.
Paired measurements show that FP8 updates are 1.90$\times$ faster
than BF16 updates.
In end-to-end GRPO across four rollout configurations,
weight-update speedups are 5.45--6.81$\times$ over veRL NCCL, with
response-token throughput gains of 5.49--6.71\%.
We also evaluate update latency and the costs of preparation and
reconfiguration during coordinated rollout scaling.

We make the following three contributions:
(1) We analyze the coupled communication, memory, and conversion costs
  of synchronous weight updates and the requirements imposed by
  rollout dynamics.
(2) We design \sys with reusable transfer planning and buffer allocation
  based on execution schedules and a persistent GPU pipeline fusing
  conversion, communication, replica forwarding, and destination writes.
(3) We evaluate \sys across models and deployment configurations, showing
  faster complete updates and higher RL throughput, and quantify
  communication-memory usage and rollout reconfiguration costs.

\section{Background and Motivation}
\label{sec:background}

\subsection{Weight Synchronization}
\label{sec:bg-critical}

In disaggregated RL, training and rollout use separate GPU pools, and weight
updates propagate newly optimized policy parameters to rollout
engines~\cite{fu2025areal,zhong2025streamrl}. We focus on \emph{synchronous,
in-place weight updates}: trainers pause optimization, and rollout engines
pause generation while new weights overwrite existing parameter
tensors~\cite{wu2025llamarl,jiang2026weightbridge}. Both pools wait until
every rollout engine has installed the complete new version. The slowest
destination determines the pause, so we target \emph{complete weight-update
latency}, including communication, conversion, and destination writes.

This update mode applies to both synchronous RL, which alternates
rollout and policy optimization~\cite{yao2023deepspeedchat,hu2025openrlhf}, and
asynchronous RL, which overlaps them and can train on trajectories
generated by earlier policy
versions~\cite{fu2025areal,wu2025llamarl,zhong2025streamrl,han2025asyncflow,sheng2026laminar,hu2026dora,li2026staleflow}.
Asynchronous \emph{weight updates} additionally overlap transfer with
computation by retaining stable source copies or staging incoming
weights separately from live
parameters~\cite{han2025asyncflow,sheng2026laminar}.
For example, WeightBridge supports sender and receiver
staging~\cite{jiang2026weightbridge}, while TensorHub exposes versioned
weight references for independent retrieval~\cite{ye2026tensorhub}.
With such overlap, transfer duration can differ from GPU stall time;
our synchronous setting directly exposes complete update latency as a
pause in both GPU pools.

Figure~\ref{fig:sync-model-size} illustrates the overhead across model
configurations using veRL NCCL. The share of the synchronization cycle
spent publishing weights is roughly twice as large for Qwen3-30B-A3B
as for Qwen3-8B. This recurring cost motivates optimizing the full
update path, from weight distribution to parameter installation.

\subsection{Limitations of Existing Solutions}
\label{sec:bg-anatomy}

Table~\ref{tab:limits} compares weight-update solutions by communication
redundancy, communication-memory footprint, precision conversion, and rollout
dynamics.

\para{Reducing redundant communication.} Training and rollout favor different
sharding configurations to fit optimizer states, activations, and KV
caches~\cite{mei2025real,wu2025llamarl}. Weights can also be replicated across
trainer ranks and within or across rollout instances. Three sources of
redundant communication arise: reconstructing full tensors
from trainer shards, delivering elements outside a receiver's shard, and
repeatedly injecting the same shard from trainers for different rollout
replicas~\cite{jiang2026weightbridge}.

\begin{figure}[!t]
  \centering
  \includegraphics[width=\columnwidth,trim=0 2.5bp 0 2.5bp,clip]{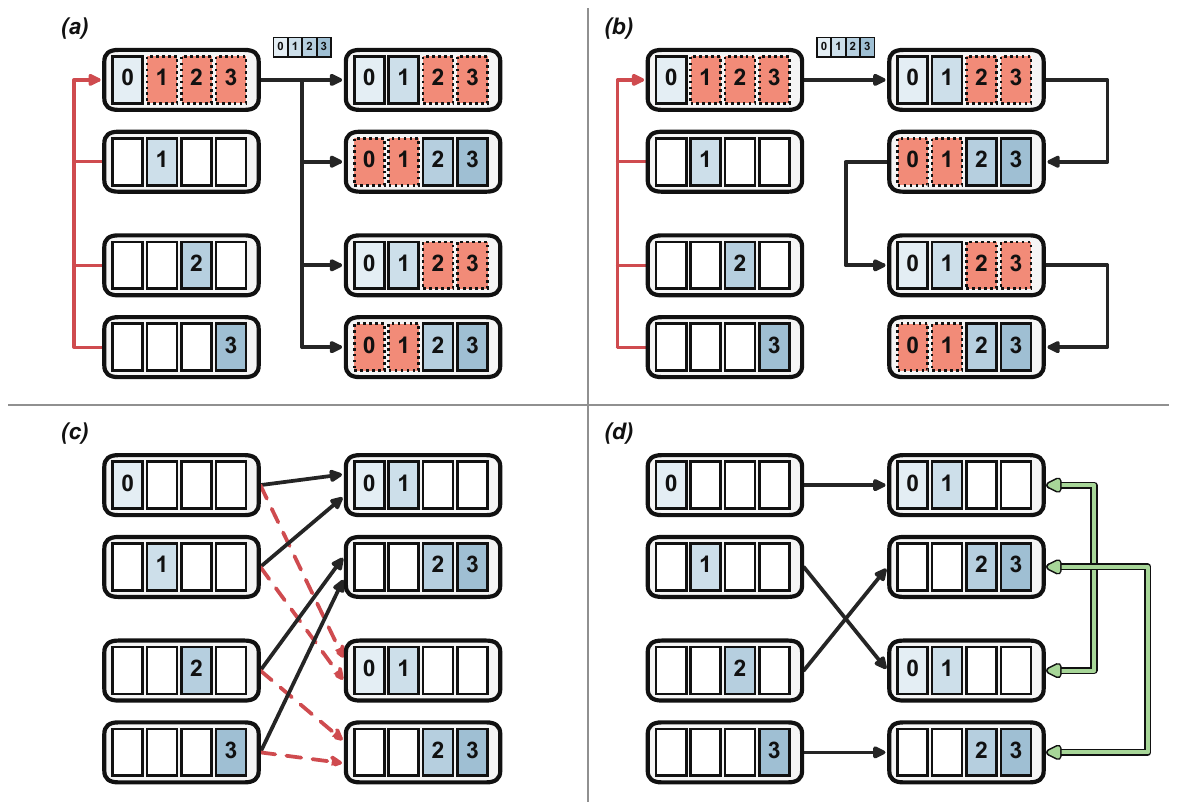}\par
  \caption{Weight-update routing patterns.}
  \label{fig:routing-comparison}
\end{figure}

Existing paths address different parts of this redundancy. Gather--broadcast
(Figure~\ref{fig:routing-comparison}a), used by veRL NCCL and slime,
reconstructs full tensors before disseminating them to rollout
workers~\cite{sheng2025hybridflow,slime2025}. Full-payload relay
(Figure~\ref{fig:routing-comparison}b), as in veRL NIXL, shares trainer
transfers through rollout forwarding, but still delivers elements that
individual workers do not need. Direct sharded P2P
(Figure~\ref{fig:routing-comparison}c), as in Awex and Miles, sends each
rollout worker only the shards it needs, but repeats trainer transfers when
equivalent destinations are served
independently~\cite{inclusionai2025awex,radixark2026miles}. WeightBridge
combines shard routing with rollout exchanges to reduce
these costs~\cite{jiang2026weightbridge}. Efficient
redistribution requires coordinating shard-specific delivery with replica
reuse. Realizing these savings also requires controlling buffering and
preparation costs along the update path.

\begin{table}[!t]
\centering
\setlength{\tabcolsep}{3pt}
\caption{Weight-update capabilities in supported layouts.
\mbox{\textbf{Redund.}}: shard delivery and replica reuse without full-tensor gathering;
\textbf{Memory}: high throughput with low communication memory;
\textbf{DType}: fused conversion;
\textbf{Dynamics}: rollout resizing.
\cmark: met; \pmark: separate conversion;
\xmark: unmet; \nmark: unreported.}
\label{tab:limits}
\vspace{0mm}
\fontsize{7.5}{8.5}\selectfont
\renewcommand{\arraystretch}{0.95}
\begin{tabular*}{\columnwidth}{@{\extracolsep{\fill}}lcccc@{}}
\toprule
System & Redund. & Memory & DType & Dynamics \\
\midrule
veRL NCCL~\cite{sheng2025hybridflow}
  & \xmark & \xmark & \pmark & \cmark \\
Awex NCCL~\cite{inclusionai2025awex}
  & \xmark & \xmark & \pmark & \xmark \\
veRL NIXL~\cite{sheng2025hybridflow}
  & \xmark & \xmark & \pmark & \cmark \\
FabricLib~\cite{licker2026fabriclib}
  & \xmark & \nmark & \pmark & \nmark \\
WeightBridge~\cite{jiang2026weightbridge}
  & \cmark & \xmark & \xmark & \xmark \\
TensorHub~\cite{ye2026tensorhub}
  & \nmark & \cmark & \pmark & \cmark \\
\midrule
\sys & \cmark & \cmark & \cmark & \cmark \\
\bottomrule
\end{tabular*}
\end{table}

\para{Communication memory footprint.} Reducing transferred bytes does not
eliminate the overhead of moving many small, scattered fragments. Batching
amortizes submission and synchronization costs, but staging and double
buffering consume GPU memory alongside model parameters and KV caches. Smaller
buffers limit this footprint but can increase transfer rounds and pipeline
bubbles~\cite{radixark2026miles}. WeightBridge reports
lower transfer latency when increasing the registered RDMA-memory budget from
1 to 2~GiB per rollout worker, with diminishing gains at larger budgets; its
default configuration uses a 4~GiB budget per rollout
worker~\cite{jiang2026weightbridge}.

\begin{figure}[!t]
  \centering
  \includegraphics[width=\columnwidth]{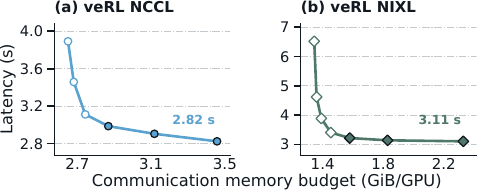}\par
  \caption{Memory--latency tradeoff (p50) for Qwen3-30B-A3B.
  Open markers denote smaller-bucket measurements.}
  \label{fig:memory-throughput}
\end{figure}

We examine this tradeoff by varying the bucket size in veRL's NCCL and NIXL
engines, each using two reusable buckets, and measuring update latency and
additional GPU-memory footprint. Figure~\ref{fig:memory-throughput} shows that
larger buckets reduce latency with diminishing returns. At the small-buffer
end, modest memory savings accompany a sharp increase in latency. In the
tested configurations, both engines reach their lowest measured latency with
several GiB of additional resident GPU memory. This measured footprint
includes communication state and retained allocator cache as well as payload
buffers. These results motivate amortizing transfer overheads within a limited
buffering budget.

\phantomsection
\label{sec:bg-conversion}

\para{Payload reduction and conversion costs.}
The destination representation introduces further work along this
path. Memory layouts differ across engines: Megatron interleaves QKV
heads by query group, while inference loaders may regroup projections.
Materializing reordered regions adds intermediate GPU-memory writes
and rereads~\cite{nvidia2025megatronbridge,hao2026rdt}.
Precision differences introduce similar costs.
Training may retain BF16 weights for numerical accuracy while rollout
uses FP8 for memory and compute efficiency, requiring quantization
and scales~\cite{wu2025llamarl,licker2026fabriclib,radixark2026miles}.
Receiver-side quantization still transfers BF16 weights.
Slime and Awex reduce the payload through source-side quantization,
but materialize quantized weights and scales before transfer, which
must subsequently read these intermediate
results~\cite{slime2025,inclusionai2025awex}.

Overlapping preparation and transfer can hide some preparation latency, but
does not remove these intermediate accesses. FabricLib overlaps parameter
preparation, including optional quantization, with RDMA; preparation and
transfer remain separate stages connected by intermediate
buffers~\cite{licker2026fabriclib}. The associated memory traffic and
coordination can erode the latency savings from sending fewer bytes. Efficient
updates must therefore coordinate conversion, transfer, and destination writes
while also reducing intermediate materialization.

\phantomsection
\label{sec:bg-elastic}

\para{Adapting to rollout dynamics.}
Rollout capacity can change during RL training: StreamRL adds rollout
nodes as generation lengthens, while ROSE borrows serving
GPUs~\cite{zhong2025streamrl,gao2026rose}.
Even with fixed per-instance layouts, these changes alter shard
destinations, replica routes, and link loads. Weight-update mechanisms
must therefore adapt their data distribution and communication state
to the active rollout set.

Existing systems support different aspects of this adaptation.
Mooncake and NIXL provide transport registration and peer
discovery~\cite{qin2025mooncake,nvidia2025nixl}, while veRL supports
rollout instance addition and removal by rebuilding the communication
topology for its NCCL and NIXL
backends~\cite{sheng2025hybridflow}.
TensorHub and Laminar allow independent adoption of coexisting
policy versions~\cite{ye2026tensorhub,sheng2026laminar}.
Transport reconfiguration and version management, however, must also
be coordinated with shard delivery, replica reuse, and representation
conversion for the changed deployment.
Under our synchronous contract, every active rollout engine must
finish installing the complete new version before generation resumes.
Efficient support for rollout dynamics therefore requires reconfiguring
the update path while meeting the preceding communication, memory,
and conversion requirements.

\section{\sys Design}
\label{sec:design}

\begin{figure}[t]
\centering
\includegraphics[width=\columnwidth,trim=0 4bp 0 5.5bp,clip]{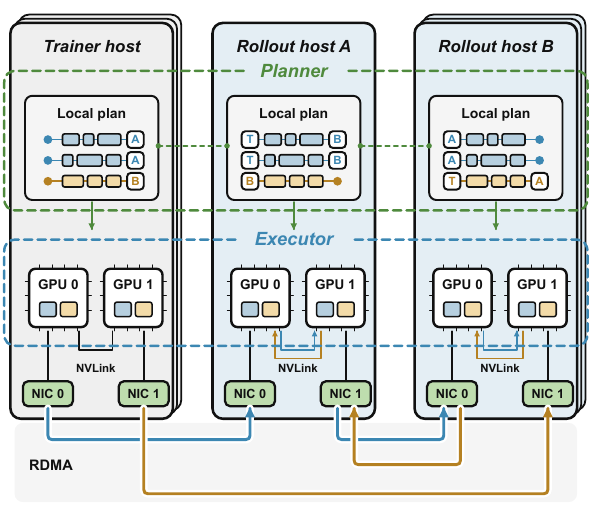}\par
\caption{\sys architecture.}
\label{fig:overview}
\end{figure}

Synchronous, in-place weight updates (\S\ref{sec:bg-critical}) offer the
following three design opportunities: \emph{Cross-tensor scheduling.} The
trainer keeps the parameter values being transferred unchanged, while rollout
engines pause generation until the weight update completes. These transfers
can therefore be reordered and batched across tensor boundaries.
\emph{Resources available during synchronous updates.} Pausing optimization
and generation reduces competition for GPU compute resources and memory
bandwidth, making these resources available for parameter conversion,
forwarding, and destination writes. \emph{Plan reuse.} Parameter values change
across weight updates. With unchanged tensor descriptors, rank placement, and
active rollout instances, execution plans and communication state can be
reused across updates.

\sys uses these opportunities to organize work across tensors while streaming
payloads through bounded buffers (Figure~\ref{fig:overview}). The
\emph{Planner} maps required parameter regions to native storage, selects
replica routes, and builds channel queues (\S\ref{sec:design-planner}). The
\emph{Executor} follows these descriptors to gather and convert source
fragments, forward payloads, and write destination tensors per update
(\S\ref{sec:design-executor}).

\subsection{\sys Planner}
\label{sec:design-planner}

Preparation combines parallel topologies, parameter metadata, native
layouts, target data types and quantization schemes, and physical rank
placement. Logical resharding, memory binding, routing, and scheduling yield
per-rank plans with complete quantization-block reads, matched FIFO
progression, and the final global completion barrier.

\subsubsection{Mapping Required Transfers}
\label{sec:design-planning}
\label{sec:design-layout}

To transfer only required data, the Planner identifies shard overlaps and
binds them to native storage. Recording quantization dependencies with these
mappings lets the Executor generate transfer payloads as it reads source
weights.

\para{Mapping shard overlaps.} A per-family resolver maps native names into a
shared logical parameter space and resolves ownership from the engine's
TP, EP, DP, CP, and PP configuration (\S\ref{sec:bg-anatomy}). Each logical parameter has
a global shape and a set of replicas, each composed of shards described by
rank, local shape, global offset, and data type. A shard is thus a
hyperrectangle in logical coordinates, independent of its native storage.
For each matched parameter, the Planner then intersects
source and destination shard hyperrectangles in logical coordinates.
Each non-empty overlap records source and destination ranks, overlap shape,
and local slices on both sides. These mappings establish logical coverage
before native-storage binding. Source-replica selection and replica
forwarding are described in \S\ref{sec:design-forwarding}.

\para{Binding native storage.} Memory-layout descriptors list contiguous
native spans in logical row-major order for fused QKV, gate--up tensors, or
stacked experts without copying them. The Planner maps overlaps to these
spans, binding tensor addresses, byte offsets, row widths, and row strides.
For example, a rollout rank's fused QKV tensor slice maps to row ranges in the
trainer's interleaved buffer. During transfer, recorded offsets and strides
guide native-tensor reads and writes, avoiding materialized reordered tensors
and a separate memory-layout conversion kernel.

\para{Planning quantized transfers.} The Planner augments storage mappings
with the source reads and destination writes required by quantization.
For BF16-to-block-FP8 transfers, computing each block's scale requires
reading all values in that block. The Planner therefore requires each
quantized source region to contain complete blocks so that scales can be
computed locally. Its descriptors record source spans and block shapes,
along with destination addresses for quantized values and scales.
The target data type and quantization scheme also determine each region's
transfer payload size, including scales and padding. These descriptors
are reused across updates; the Executor computes scales and quantized
values from the current weights during transfer.

\subsubsection{Replica Route Planning}
\label{sec:design-forwarding}

\begin{figure}[t]
\centering
\includegraphics[width=0.76\columnwidth,trim=0 1.5bp 0 1.5bp,clip]{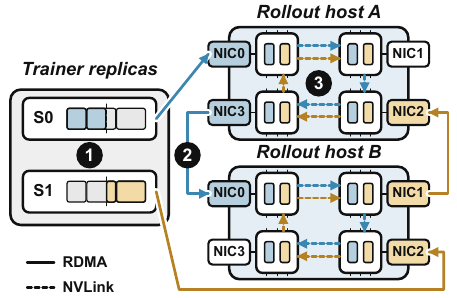}\par
\caption{Replica route planning.}
\label{fig:ring}
\end{figure}

Weights may be replicated across trainer ranks,
across ranks within a rollout instance, or across
independent rollout instances.
The Planner uses shard offsets and shapes to identify logical parameter
regions, recording all trainer ranks that hold each region and all rollout
ranks that require it (Figure~\ref{fig:ring}).
Using these rank sets, the Planner selects senders and forwarding paths to reuse
transmissions while distributing communication across ranks and NICs.

Given trainer ranks holding copies of the
same parameter regions, the Planner concatenates those regions into a virtual
byte stream and assigns each rank a disjoint byte range to send
(Figure~\ref{fig:ring}, \ding{172}). This balances source data across trainer
ranks without splitting every parameter across all replicas, even with
unequal parameter sizes.

For parameter slice $j$, let $W_j$ be its payload size, including scales and
padding, and $r_j$ its destination count. Direct P2P sends $\sum_j r_jW_j$
bytes from trainers. \sys instead forms a relay chain through all $r_j$
destinations, rooted at the selected source (\ding{173}). Trainer traffic falls
to $W=\sum_j W_j$, reusing source-side transformations and payloads within
and across rollout instances. Relays write received values to parameter
tensors and forward the same payload. Chunk
pipelining~\cite{thakur2005mpich,chan2007collective,zhuang2023optimizing}
amortizes hop latency.

Each relay chain serves all required ranks
within a host before forwarding the slice to another host, so each slice
enters each required host only once (\ding{174}). To prevent trainer traffic
from concentrating on one rollout host, the Planner selects each chain's
first host by accumulated source-byte load. Within each host, fixed entry
and exit ranks would concentrate inter-host traffic on their associated
NICs. The Planner therefore rotates the local rank order across chains,
distributing receive and forwarding work across ranks. Combined with
NUMA-aware NIC assignment, this spreads traffic across local NICs while
preserving GPU--NIC locality.

\subsubsection{Cross-Tensor Execution Planning}
\label{sec:design-scheduling}

Execution planning takes transfer descriptors of native tensor
addresses, byte offsets and strides, quantization metadata, and payload sizes
(\S\ref{sec:design-planning}), plus selected senders and ordered
relay chains (\S\ref{sec:design-forwarding}). These descriptors specify
uneven, potentially small payload spans. To amortize per-operation
overhead while limiting communication memory, the Planner groups descriptors
across tensor boundaries and builds demand-based channel queues.
Work batches describe data entering FIFO slots during
execution; full batch payloads need not be materialized in advance.

\begin{figure}[t]
\centering
\includegraphics[width=0.85\columnwidth,trim=0 2.5bp 0 3.5bp,clip]{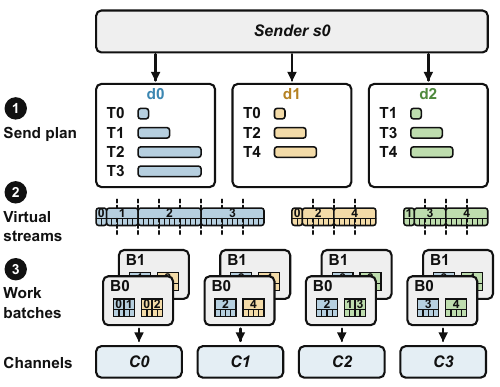}\par
\caption{Sender-side execution planning.}
\label{fig:planning}
\end{figure}

For each rank, the Planner groups transfer
descriptors by relay chain and communication peers (Figure~\ref{fig:planning},
\ding{172}). Within each group, it concatenates descriptors across
tensors into a virtual byte stream without copying tensor
data (\ding{173}).
Within the preparation-time resource limits (\S\ref{sec:design-resources}),
the Planner selects channel counts and partitions each stream into
approximately equal, aligned ranges across channels, then divides each range
into bounded \emph{work items}.
Each item records its communication peers, byte offset within the stream,
and byte count. Its fragment descriptors identify the native addresses and
quantization metadata for those bytes. Items may span tensors, and tensors
may span channels.

Within each channel, the Planner groups work items into \emph{work batches} of
up to eight items with distinct communication peers (\ding{174}). It
coordinates stream partitions and work ordering across the relay chain to
match sends, receives, and forwarding operations. The per-channel batch queues
and fragment descriptors form each rank's execution plan.

\subsubsection{Plan Reuse and Rollout Scaling}
\label{sec:design-membership}

After address and route binding, collective preparation negotiates channel
ceilings, creates communication state and registered FIFO storage
(\S\ref{sec:design-resources}), and uploads the execution plan.
It can run explicitly before an update or
lazily on first use. Explicit preparation touches no model data and runs no
transfer kernel. Subsequent updates reuse cached descriptors and queues;
changes to bound addresses, layouts, or the active rollout instances require
new preparation.

\sys handles coordinated rollout additions and suffix removals through
plan replacement, with trainer size and per-instance topology fixed.
The coordinator defers replacement until the current weight update completes.
Ranks in the updated participant set prepare new routes, channel queues,
and communication state, while departing ranks drain outstanding work and
release old communication resources. Preparation preserves the retained
engines' parameter storage and last complete weights. The coordinator
activates the new plan only after all ranks in the updated set acknowledge
preparation and all departing ranks acknowledge retirement. The next weight
update uses this prepared plan to synchronize the updated rollout set
(\S\ref{sec:eval-elastic}).

\subsection{\sys Executor}
\label{sec:design-executor}
\label{sec:design-execution}

The Planner gives each rank channel queues of work items and fragment
descriptors specifying native tensor accesses and quantization metadata. The
Executor uses these queues to prepare and transfer payloads, forward received
data when required, and write values into destination tensors.

\subsubsection{Resource Provisioning}
\label{sec:design-resources}

Before a transfer kernel runs, collective preparation provisions the Executor's
channels and FIFO storage, reused with the execution plan
(\S\ref{sec:design-membership}) rather than allocated per update.

\begin{figure}[t]
  \centering
  \includegraphics[width=\columnwidth,trim=0 1.5bp 0 0,clip]{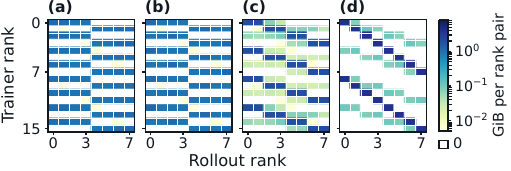}\par
  \caption{Trainer-to-rollout payload demand: Qwen3-30B-A3B BF16,
  16 trainer/8 rollout GPUs.
  Trainers: PP4/TP2/CP2/EP4 in (a,c), DP2/TP4/CP2/EP8 in (b,d);
  rollouts: two TP4 instances in (a,b), one TP8/EP8 instance in (c,d).}
  \label{fig:pair-demand}
\end{figure}

Figure~\ref{fig:pair-demand}(a--d) has
72/72/88/44 active rank pairs and maximum-to-mean payload ratios of
2.23/2.30/4.02/7.85, averaging over all rank pairs including zero-demand pairs.
This skew motivates assigning more channels to peers with larger payloads.
For inter-host transfers, the setup-time allocator derives per-peer channel
ceilings from payload demand, subject to CTA, connection, and FIFO capacity
limits. Each pair adopts the smaller ceiling supported by its endpoints;
relay chains use a common ceiling to ensure consistent channel partitions
across participants. The Planner selects the actual channel counts within
these ceilings when building queues (\S\ref{sec:design-scheduling}).
Appendix~\ref{sec:allocation-details} provides the capacity model and the
allocation algorithm.

Once the channel queues are built, the Planner identifies
the peer--channel pairs that require payload storage, including relay buffers.
Preparation allocates FIFO storage only for those pairs. Each required pair
retains its own FIFO with unchanged chunk size and depth. Eliminating unused
FIFO reservations reduces the communication memory footprint while preserving
scheduled concurrency (\S\ref{sec:eval-memory}).
Appendix~\ref{sec:compact-evidence} gives the corresponding storage-capacity model.

\subsubsection{Fused Streaming Execution}

The Executor fuses preparation and transfer in one persistent GPU kernel per
rank.
Each CTA processes one channel's batches in order, assigning disjoint warp
groups to concurrent work items. CTA barriers separate batches; each item
uses fragment descriptors for native accesses and follows its planned FIFO
sequence.

\begin{figure}[t]
\centering
\includegraphics[width=\columnwidth,trim=0 0 0 0.75bp,clip]{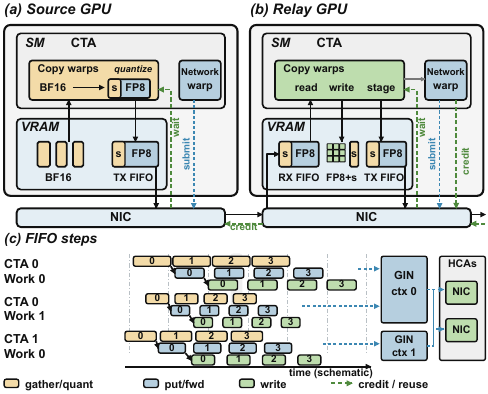}\par
\caption{Warp-cooperative execution. (a--b) Copy and network warps in one
work item's warp group. (c) Work items in CTA 0 and CTA 1;
numbers 0--3 identify FIFO steps within each item.}
\label{fig:pipeline}
\end{figure}

\para{Pipelining and buffer reuse.} Each work item streams through fixed-depth
FIFOs in payload \emph{chunks}, one per slot. Reusing slots at this granularity,
rather than buffering rounds as in WeightBridge or prepared tensors as in
FabricLib~\cite{jiang2026weightbridge,licker2026fabriclib}, limits the
communication memory footprint without materializing full work batches
(\S\ref{sec:design-scheduling}).

GIN work items with at least three warps separate copying from network
submission. For sends, \emph{copy warps} gather and optionally quantize native fragments into FIFO
slots. After a copy barrier, a network warp submits completed chunks while
copy warps prepare the next slot (Figure~\ref{fig:pipeline}); NVLink (LSA)
accesses peer-visible FIFOs directly. Consumers wait for \emph{readiness},
published after data becomes
visible; producers reuse slots only after \emph{credits} confirm all readers
have finished.

Direct relays hold receive slots until local parameter writes and outgoing
reads finish. Staged GIN-to-GIN relays instead copy data into destination
tensors and a separate send slot, then publish a completion watermark and
release the receive slot. A network warp waits for the
watermark before submission while copy warps process subsequent chunks without
waiting for network submission. When the send FIFO is full, copy warps wait
for downstream credits before reusing a slot, avoiding overwrites of chunks
still in flight.

\para{Fused block quantization.} For BF16-to-FP8 transfers, the payload
contains both quantized values and their block scales. Each $b_r\times b_c$
tile is transmitted as a record: a 16-byte header containing an FP32 scale
and padding, followed by
$b_rb_c$ one-byte FP8 (E4M3) values. For $n$ source elements in complete
tiles, the transfer payload is $n(1+16/(b_rb_c))$ bytes, about half the
size of the BF16 source.

To produce a record, a warp reads a full BF16 tile through its
fragment descriptor, computes the block scale, and writes the scale and
FP8 values directly into FIFO storage. No separate quantized tensor is
materialized. For records crossing channel, work-item, or chunk boundaries,
each invocation reads the full tile but produces only its assigned
bytes, potentially recomputing scales. Relays
forward these bytes unchanged. Receivers extract and write FP8 values and
scales into their respective destination tensors.

\section{Implementation}
\label{sec:implementation}

The Planner uses approximately 4.5k lines of Python for parameter
metadata, logical planning, and rollout scaling, and 2.3k lines of C++ for
transport preparation and plan construction. The Executor uses
1.4k lines of CUDA. We integrate \sys with Megatron for trainer-side parameter
descriptors, vLLM for direct writes into live weight and scale tensors, and
veRL as a weight-update backend.
We use NCCL Device API peer-memory load/store access (LSA) for NVLink
transfers and GPU-initiated networking (GIN) for inter-host RDMA. On our
testbed, GIN's GPUDirect Async Kernel-Initiated (GDAKI) backend submits RDMA
directly from GPU threads to reduce CPU-proxy dispatch and CPU--GPU
coordination~\cite{hamidouche2025gin}. Multiple GIN contexts and connections
reduce submission-queue contention.
We bind ranks to nearby active NIC ports and pin worker threads to CPUs in
the corresponding GPU NUMA domains to limit cross-NUMA traffic. GPU placement
and port assignment spread transfers across NICs to reduce contention.

\begin{figure*}[!t]
  \centering
  \includegraphics[width=\textwidth,trim=0 1bp 0 0,clip]{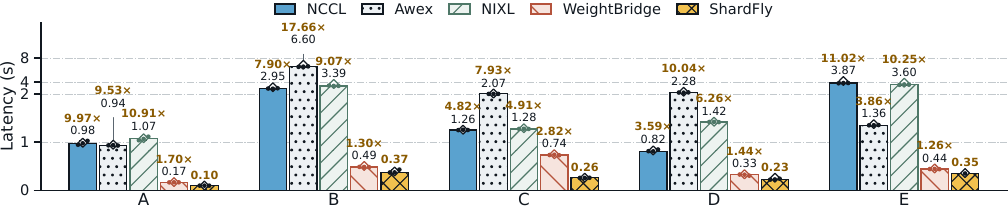}\par
  \caption{Complete BF16 update latency across model configurations. Dots show independent
  launch p50s; bars their median, whiskers their range, and hollow diamonds
  the median launch p95. Gold ratios above each baseline give its p50
  divided by ShardFly's p50.}
  \label{fig:topology}
\end{figure*}

\section{Evaluation}
\label{sec:evaluation}

We evaluate the system to answer three questions:
\begin{enumerate}[label=\textbf{Q\arabic*.},leftmargin=*]
\item How efficient is \sys when performing weight updates across models,
  deployment configurations, replica counts, data types, and
  communication-memory footprints (\S\ref{sec:eval-performance})?
\item Can \sys support rollout scaling, and at what
  reconfiguration cost
  (\S\ref{sec:eval-elastic})?
\item How much does \sys improve end-to-end RL training performance
  (\S\ref{sec:eval-e2e})?
\end{enumerate}

\subsection{Experiment Setup}
\label{sec:eval-setup}

\para{Testbed and workloads.} We use four hosts with 32 GPUs in total.
Each host has eight NVIDIA H20-3e GPUs (141\,GB each) in one NVLink
domain and four 400\,Gb/s RoCE links, providing a nominal one-way
aggregate network bandwidth of 200\,GB/s per host (50\,GB/s per link).
Training and rollout occupy separate hosts. We use Megatron for training
and vLLM 0.27 for rollout, with PyTorch 2.13, CUDA 13.2, and Ray 2.58.
Standalone experiments measure weight updates without optimizer steps;
end-to-end RL experiments run GRPO with synchronous weight updates.
Appendix~\ref{sec:reproducibility} gives measurement details, including the
correctness-validation harness (\S\ref{sec:correctness-validation}).

\para{Models and metrics.} We test dense Qwen3-8B and four MoE
models: Qwen3-30B-A3B, GPT-OSS-20B, Moonlight-16B-A3B, and GLM-4.7-Flash.
Our primary metric is complete weight-update latency, measured as wall-clock
time for the full update workflow, including weight export, communication,
destination loading, global synchronization, and control-message exchanges.
Process/model startup and post-update correctness checks are excluded.
For end-to-end RL, we report response tokens consumed per second of the
complete training loop and blocked rollout GPU time.

\para{Baselines.} We compare four standalone weight-update backends:
\mbox{(i) \emph{Gather--Broadcast}:} veRL's NCCL
checkpoint engine~\cite{sheng2025hybridflow} reconstructs logical weights
and broadcasts them in buckets; each receiver loads its required portion.
\mbox{(ii) \emph{Direct P2P}:} Awex's NCCL transport~\cite{inclusionai2025awex}
sends source--destination shard overlaps directly, serving each rollout
replica from the trainer without receiver relays.
\mbox{(iii) \emph{Pull--Relay}:} veRL's NIXL
engine~\cite{sheng2025hybridflow} pulls exported weight buckets
from a predecessor and exposes them to the next receiver in a chain.
\mbox{(iv) \emph{WeightBridge}:} WeightBridge~\cite{jiang2026weightbridge}
transfers required shards with replica deduplication and reuses cross-tensor
plans to pipeline packing, communication, and assembly across rounds.

\subsection{Weight-Update Performance}
\label{sec:eval-performance}

\begin{figure}[!t]
  \centering
  \includegraphics[width=\columnwidth,trim=0 1.75bp 0 1bp,clip]{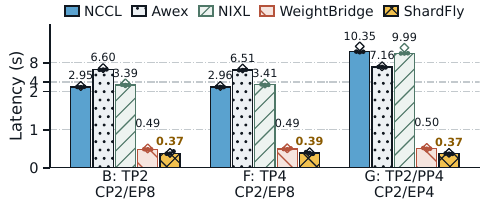}\par
  \caption{BF16 update latency across trainer configurations.
  Markers follow Figure~\ref{fig:topology}.}
  \label{fig:trainer-layouts}
\end{figure}

For Figures~\ref{fig:topology} and~\ref{fig:trainer-layouts}, launches
measure 30 updates after five warmups; we
report median launch p50s and p95s. veRL NCCL, NIXL, and WeightBridge use
256\,MiB buckets, 256\,MiB buckets, and the default 2\,GiB round cap,
respectively, for moderate GPU-memory overhead and latency near measured
minima (Figure~\ref{fig:memory-throughput}).

\para{Across models.}\phantomsection\label{sec:eval-topo}
Across configurations A--E in Table~\ref{tab:publication-configs},
\sys completes an update in
0.098--0.374\,s, achieving 1.26--17.66$\times$ speedups across the four
baselines (Figure~\ref{fig:topology}). WeightBridge, which already deduplicates shards,
batches tensors, and reuses plans, is the fastest baseline across A--E
in these separately collected runs.

\begin{table}[!t]
\centering
\caption{BF16 configurations for Figures~\ref{fig:topology}
and~\ref{fig:trainer-layouts}.}
\label{tab:publication-configs}
\vspace{1mm}
\begingroup
\fontsize{8}{9.4}\selectfont
\setlength{\tabcolsep}{1.5pt}
\renewcommand{\arraystretch}{1.08}
\begin{tabular*}{\columnwidth}{@{\extracolsep{\fill}}llrccc@{}}
\toprule
\textbf{ID} & \textbf{Model} & \shortstack{\textbf{Weights}\\(GB)} & \shortstack{\textbf{Hosts}\\T/R} & \shortstack{\textbf{Trainer}\\{\fontsize{6.5}{7.5}\selectfont TP/PP/CP/EP/DP}} & \shortstack{\textbf{Rollout}\\{\fontsize{6.5}{7.5}\selectfont $n\times$(TP/DP/EP)}} \\
\midrule
A & Qwen3-8B & 16.4 & 2/1 & 2/2/1/1/4 & $4\times$(2/1/1) \\
B & Qwen3-30B-A3B & 61.1 & 2/2 & 2/1/2/8/4 & $4\times$(4/1/1) \\
C & GPT-OSS-20B & 41.8 & 2/1 & 4/1/1/8/4 & $2\times$(4/1/1) \\
D & Moonlight-16B-A3B & 31.9 & 2/2 & 4/1/1/8/4 & $4\times$(4/1/1) \\
E & GLM-4.7-Flash & 62.5 & 2/1 & 2/2/2/8/2 & $1\times$(1/8/8) \\
\midrule
F & Qwen3-30B-A3B & 61.1 & 2/2 & 4/1/2/8/2 & $4\times$(4/1/1) \\
G & Qwen3-30B-A3B & 61.1 & 2/2 & 2/4/2/4/1 & $4\times$(4/1/1) \\
\bottomrule
\end{tabular*}

\endgroup
\end{table}

\para{Across trainer configurations.} Figure~\ref{fig:trainer-layouts}
examines how trainer sharding affects complete updates for a fixed
rollout demand. We fix Qwen3-30B-A3B, two trainer hosts, and four TP4
rollout engines on two rollout hosts, while varying trainer parallelism
(B/F/G). All methods have similar latency between B and F: changing
TP2 to TP4 leaves PP1, EP8, and expert TP1 unchanged, preserving the
expert-weight partitioning and the destination weights to be updated.
Changing from B to G has a larger effect on veRL NCCL and NIXL: their update
latencies rise from 2.95/3.39\,s to 10.35/9.99\,s. Both paths reconstruct full
tensors through the same export routine, which additionally broadcasts weights
across pipeline stages when PP$>1$.
In contrast, WeightBridge remains within 0.488--0.503\,s, while \sys
stays within 0.371--0.392\,s across all three configurations, a further
1.26--1.36$\times$ speedup. Neither reconstructs full tensors.

\begin{figure}[!t]
  \centering
  \includegraphics[width=\columnwidth,trim=0 1bp 0 2bp,clip]{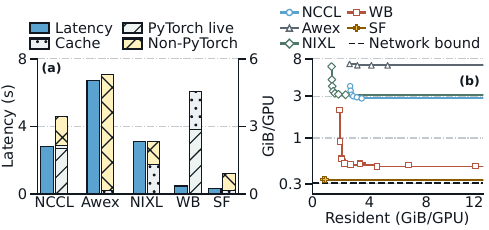}\par
  \caption{Communication memory and p50 latency (WB: WeightBridge; SF: \sys).
  (a) Fastest measured settings. (b) Best measured latency at each
  resident threshold.
  Dashed: 305\,ms network-only reference.
  Details in Appendix~\ref{sec:memory-measurement}.}
  \label{fig:memory-comparison}
\end{figure}

\para{Communication memory footprint.}\phantomsection\label{sec:eval-memory}
Figure~\ref{fig:memory-comparison} measures communication memory for
Qwen3-30B-A3B BF16 updates on two hosts: additional steady resident
memory on rollout GPUs after model loading, including allocator cache and
library-managed allocations but excluding validation copies. In panel (a), \sys
updates in 335\,ms using 0.91\,GiB/GPU, versus 2.33--5.28\,GiB/GPU
for the baselines. WeightBridge also achieves sub-second latency, but its
fastest measured setting uses 4.54\,GiB/GPU for a 473\,ms update.
Jointly reducing its round and logical-tensor caps to 32 and 8\,MiB lowers
the footprint to 2.05 and 1.89\,GiB/GPU, while latency rises to 580 and
2083\,ms, respectively.
These separately collected system measurements show that \sys combines
smaller resident memory with shorter complete updates.
Bounded streaming and compact FIFO allocation
(\S\ref{sec:design-resources}) limit \sys's footprint.
Reducing \sys's FIFO depth from 16 to 8 (Figure~\ref{fig:memory-comparison}(b))
lowers its footprint from 0.91 to 0.79\,GiB/GPU, while update p50 stays
near 335--336\,ms. Only \sys has measured configurations below
1\,GiB/GPU resident overhead.

\begin{figure}[!t]
  \centering
  \includegraphics[width=\columnwidth,trim=0 0.5bp 0 0,clip]{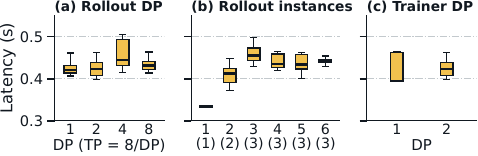}\par
  \caption{Qwen3-30B-A3B BF16 replica scaling. (a) Rollout uses EP8.
  (b) Tick labels give rollout instance counts, with host counts in parentheses.
  (c) Trainer DP varies for both dense and expert weights, with two
  TP4/DP2/EP8 rollout instances.
  Boxes pool three launches: p25--p75, p50 lines, and p5--p95 whiskers.}
  \label{fig:fanout}
\end{figure}

\para{Replica scaling.}\phantomsection\label{sec:eval-fanout}
Within two eight-GPU rollout instances, increasing DP from 1
to 8 while reducing TP from 8 to 1 keeps update p50 within 420.0--444.6\,ms
(Figure~\ref{fig:fanout}(a)). Replica relays (\S\ref{sec:design-forwarding})
keep total trainer payload at 61.064\,GB in both rollout sweeps.
Across independent TP4
instances, p50 is 333.8\,ms for one instance and 412.6--455.4\,ms for two--six
instances on up to three hosts. With three fixed rollout hosts, increasing
instances from three to six keeps p50 within 434.0--455.4\,ms
(Figure~\ref{fig:fanout}(b)); host-consecutive routing avoids repeated
inter-host transfers for additional instances on the same host. Doubling
trainer DP halves the payload per trainer host through the Planner's
source-replica selection, with unchanged rollout demand.
P50 stays similar at 395.1/423.1\,ms (Figure~\ref{fig:fanout}(c)).

\begin{figure}[!t]
  \centering
  \includegraphics[width=\columnwidth]{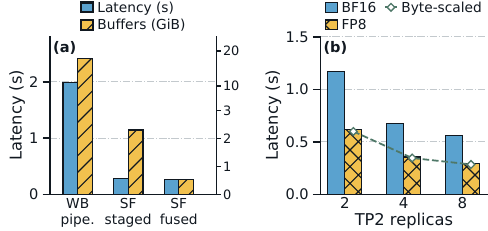}\par
  \caption{(a) FP8 update p50 (left axis) and maximum tracked extra source
  buffers per GPU (right): non-parameter PyTorch allocations plus SF native
  FIFO and quantization buffers, excluding cache and untracked runtime state.
  (b) Paired BF16/FP8 p50; dashed: BF16 latency scaled by the payload ratio.}
  \label{fig:fp8}
\end{figure}

\para{Fused quantization and transfer.}\phantomsection\label{sec:eval-fp8}
We evaluate Qwen3-30B-A3B with 16 trainer GPUs and four TP2 rollout engines,
using E4M3 FP8 with 128$\times$128 blocks. Our pipelined WeightBridge extension
adds source-side quantization before native packing, gathering incomplete
tensors within trainer TP groups and producing FP8 values and FP32 scales.
Each tensor is prepared
once, before the first round using its values or scales; existing double
buffers allow later-round preparation to overlap earlier RDMA. Native routing
and receiver consumption are unchanged, while gather and quantized buffers
remain allocated. To isolate fusion, we also construct a staged \sys control:
a separate kernel materializes deduplicated FP8 source views and scales,
which the transport then copies into its FIFO. The quantizer, wire records,
GIN backend, routes, FIFO geometry, and destination writes are unchanged.
Figure~\ref{fig:fp8}(a) shows that the WeightBridge extension takes
1.982\,s versus 0.257\,s for fused \sys, with tracked extra source allocations
of 17.79 versus 0.52\,GiB/GPU. Compared with staged \sys, fusion is 1.085$\times$ faster
and removes up to 1.779\,GiB/GPU of quantization buffers, reducing tracked
source allocations from 2.30 to 0.52\,GiB/GPU. Matched BF16/FP8
runs at two, four, and eight TP2 replicas (Figure~\ref{fig:fp8}(b)) further
show payload falling to 51.1\% and latency to 52.6--52.7\% of BF16,
indicating nearly byte-proportional latency savings.

\subsection{Elastic Rollout}
\label{sec:eval-elastic}

\begin{figure}[!t]
  \centering
  \includegraphics[width=\columnwidth,trim=0 0 0 2.25bp,clip]{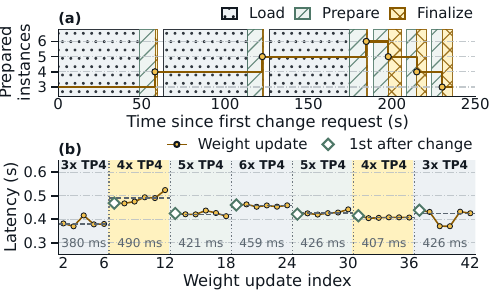}\par
  \caption{Elastic rollout scaling: (a) membership timeline;
  (b) per-update latency. Dashed lines and labels show steady p50s.}
  \label{fig:elastic}
\end{figure}

Figure~\ref{fig:elastic} evaluates \sys's ability to maintain efficient weight
updates during dynamic rollout scaling. The design in
\S\ref{sec:design-membership} preserves retained instances' parameter storage,
limiting model loading to newly added instances. New-instance loading
dominates growth, while shrinkage pays for preparation and release of
departing instances (Figure~\ref{fig:elastic}(a)). Both transitions prepare
the updated communication state and execution plan before activation, letting
the next update directly use the new participant set. Each first post-change
update is within 4.5\% of that configuration's steady p50
(Figure~\ref{fig:elastic}(b)), supporting the separation of membership
preparation from repeated updates: resizing has a transition cost, but the
prepared plan delivers near-steady update latency immediately upon activation.

\subsection{End-to-End RL Performance}
\label{sec:eval-e2e}

We evaluate whether faster weight updates improve end-to-end RL throughput. We
integrate \sys as a veRL weight-update backend and train Qwen3-30B-A3B with
fully asynchronous GRPO on GSM8K, using separate 16-GPU H20 pools for training
and rollout. The trainer publishes BF16 weights after every optimizer update.
Appendix~\ref{sec:rl-experiment-details} gives the configurations and
measurement details.
Compared with veRL's native NCCL backend, \sys reduces complete
weight-update latency from 6.89--7.20\,s to
1.03--1.26\,s across the four configurations (Table~\ref{tab:rl-e2e}),
including rollout pause, request cancellation/draining, and resume.
Even in asynchronous GRPO, weight synchronization pauses the entire
rollout pool. Shorter updates reduce this generation-blocked time,
measured as pause duration times the 16 rollout GPUs, from 109--113 to
13--19 GPU-seconds per update.
These shorter pauses translate into
response-token throughput gains of 5.49--6.71\% across all four
configurations. The complete RL loop also includes
generation and optimization, yielding a smaller but consistent throughput gain.

\begin{table}[!t]
\centering
\caption{End-to-end RL results across configurations R1--R4.
Throughput counts response tokens.}
\label{tab:rl-e2e}
\begingroup
\fontsize{7.2}{8.5}\selectfont
\begin{tabular*}{\columnwidth}{@{\extracolsep{\fill}}lrrrrrr@{}}
\toprule
 & \multicolumn{2}{c}{Throughput (token/s)} & \multicolumn{2}{c}{Update (s)} & \multicolumn{2}{c}{Stall (GPU$\cdot$s)} \\
Layout & NCCL & Ours & NCCL & Ours & NCCL & Ours \\
\midrule
R1 & 15,871 & 16,936 & 6.89 & 1.26 & 111.59 & 19.19 \\
R2 & 15,183 & 16,016 & 7.20 & 1.07 & 112.74 & 13.34 \\
R3 & 14,953 & 15,846 & 7.03 & 1.03 & 109.37 & 12.77 \\
R4 & 15,071 & 16,048 & 6.90 & 1.14 & 108.72 & 14.66 \\
\bottomrule
\end{tabular*}

\endgroup
\end{table}
\par

\section{Other Related Work}
\label{sec:related-work}

\para{RL post-training systems.} RL systems optimize placement, scheduling,
and execution across training and
rollout~\cite{zhong2025rlhfuse,wang2025roll,gao2026rollpacker,he2026rhymerl,yu2026rlinf,qin2026seer}.
These optimizations complement \sys's faster weight updates.

\para{Model loading and resharding.} Serverless systems multicast replicas
with identical layouts through chains or binomial
pipelines~\cite{zhang2025blitzscale,yu2025lambdascale}. Checkpoint resharding
maps weights between partitioning
configurations~\cite{zheng2022alpa,zhuang2023optimizing,wan2025bytecheckpoint,lian2025ucp}.
\sys targets repeated updates between live trainer and rollout parameters.

\para{Collectives and GPU-initiated communication.} Synthesized
collectives~\cite{shah2023taccl,cowan2023msccl,liu2024teccl}
target homogeneous buffers. NCCL M2N provides layout-aware resharding of a
logical tensor and has been integrated into full-model weight
updates~\cite{kandadi2026ncclm2n}. Its PIPE transport already supports bounded
chunks with device-side or host-driven RMA progress. GPU-initiated
primitives~\cite{potluri2017nvshmem,hamidouche2025gin,hwang2026mscclpp}
support MoE all-to-all~\cite{zhao2025deepep,licker2025pplxkernels,uccl2025ep}
and compute--communication
fusion~\cite{zheng2025tilelink,zhang2025comet,sul2026parallelkittens}.
\sys jointly organizes cross-tensor execution, native-memory access,
quantization, replica forwarding, and buffer reuse for a complete weight
update.

\para{Sparse and delta weight transfer.} Sparse and delta transfers reduce
payloads by sending selected or changed
weights~\cite{hu2026sparserlsync,miahi2026pulse,peng2026arealdte}.
We evaluate full-weight updates without these complementary techniques.

\section{Conclusion}
\label{sec:conclusion}

\sys combines reusable transfer plans and fused GPU streaming for
synchronous RL weight updates. Evaluation shows lower weight-update latency
across models and deployments and higher end-to-end RL training throughput.

\bibliographystyle{abbrv}
{\scriptsize\bibliography{references}}

\clearpage
\appendix
\section{Channel Allocation Details}
\label{sec:allocation-details}

\para{Demand-aware resource allocation.} This appendix gives the capacity model and the setup-time channel allocator
summarized in \S\ref{sec:design-resources}. The model accounts for
shared bandwidth, conversion, and staging; the implementation uses the
payload-based heuristic below during preparation.
FIFO chunks contain $\sigma$ bytes and each FIFO has $D$ slots.
A stream $r$ is a direct transfer or an entire relay chain with transfer
payload $W_r$. Expanding its physical hops gives the traffic $Q_\ell$ on
each shared resource $\ell$, including NIC transmit/receive, NVLink, and GPU
memory. Memory traffic counts native reads, staging, and destination writes.
With usable capacities $C_\ell$,
$T_0=\max_\ell Q_\ell/C_\ell$ is a shared-resource lower bound on transfer
duration. Ranks on the same NIC share one capacity
constraint across all GIN connections.

For fixed FIFO settings, let $p_r$ denote the slowest stage's per-channel
service rate in wire bytes/s, including source conversion, destination
writes, and relay work. Let $\tau_r$ bound credit turnaround, including
downstream consumption. An initial stream-rate model is
\begin{equation}
 q_r(c)=c\min\!\left\{p_r,\frac{D\sigma}{\tau_r}\right\}.
 \label{eq:channel-rate}
\end{equation}
Channels parallelize preparation and keep bytes in flight between credit
returns, subject to shared capacities.

For fixed routes and registration geometry $g$, the decision variables are
the modeled completion time $T$ and channel reservations $c_r$.
Let $\mathcal C_r$ contain the legal powers of two supported by every rank on
stream $r$, $\delta_r$ its pipeline fill/drain delay, and $A_k$ rank $k$'s
CTA allowance. Let $M_k(g)$ be its registered-window footprint and
$\overline M_k$ its memory allowance. The resulting reservation problem is
\begin{equation}
\begin{aligned}
 \underset{T,\{c_r\}}{\operatorname{minimize}}\quad & T \\
 \text{subject to}\quad
 & Q_\ell/C_\ell \le T && \forall\ell, \\
 & \delta_r+W_r/q_r(c_r)\le T && \forall r, \\
 & \sum_{r:\,k\in r}c_r\le A_k && \forall k, \\
 & M_k(g)\le\overline M_k && \forall k, \\
 & c_r\in\mathcal C_r && \forall r.
\end{aligned}
\label{eq:allocation-problem}
\end{equation}
Here $k\in r$ includes senders, receivers, and relays, which reserve the same
channel count for a chain. Execution shares channels through batches and queues
(\S\ref{sec:design-scheduling}).

For a candidate $T\ge T_0$, the minimum reservation needed by each stream is
\begin{equation}
 c_r(T)=\min\!\left\{c\in\mathcal C_r:\;
 \delta_r+W_r/q_r(c)\le T\right\},
 \label{eq:channel-allocation}
\end{equation}
An empty feasible set rejects $T$. For fixed rates and geometry, these counts
yield a monotone feasibility test over finite completion thresholds.
\sys uses this principle in a setup-time heuristic, fixing window
geometry and balancing payload within transport- and data-type-dependent
concurrency budgets. This heuristic uses payload estimates and fixed budgets,
without fitting stage service rates online.

\phantomsection\label{sec:compact-evidence}
Window memory separately constrains feasibility. \sys allocates payload
storage only for peer--channel pairs used by the lowered schedule: at both
endpoints for GIN and at the producer for LSA. With $U_{\max}$ the maximum
required slot count across ranks, symmetric payload capacity is
$U_{\max}D_{\mathrm{win}}S_{\mathrm{slot}}$, where $D_{\mathrm{win}}$ is
the window depth and $S_{\mathrm{slot}}$ the physical slot size.
The registered footprint $M_k(g)$ additionally includes control state for
the $K$ registered channels and alignment padding. Changes to the required
storage mapping trigger new preparation.

\para{Demand-balanced channel allocation.} Within this budget, \sys assigns
more GIN channels to peers with larger planned payloads. At rank $k$, let $\mathcal P_k$ be
its active GIN peers, $B_{kp}$ the accumulated payload for peer $p$, and
$x_{kp}$ the peer's channel ceiling. Approximating channels within a
transport class by a common service rate $v_k$ gives a completion-time term
$B_{kp}/(v_k x_{kp})$. The common rate cancels when comparing peers, leaving
$B_{kp}/x_{kp}$ as the allocation priority. In the continuous relaxation
without per-peer bounds, equalizing this pressure assigns channels in
proportion to payload.

The implementation bounds this proportional allocation by available
transport concurrency and FIFO demand. With $J_k$ GIN connections and $K$
registered channels, a transport- and data-type-dependent factor $f$ sets the
nominal budget $a_k=\min(K,fJ_k)$. The factor provides concurrency for data
preparation and communication; the larger budgets for direct FP8 fanout
accommodate the source conversion work described above
(\S\ref{sec:implementation}). For a small payload, allocating many channels
would divide little work across multiple FIFOs. We therefore use the number
of FIFO windows needed to hold the peer's payload, $\lceil B_{kp}/(D\sigma)\rceil$,
to limit its demand, together with a connection-based cap $L_k$.
Writing $\lceil z\rceil_2$ for rounding up to a power of two, the desired
peer ceiling is
\begin{equation}
\begin{aligned}
 L_k &= \min\{K,\lceil 2J_k\rceil_2\}, \\
 h_{kp} &= \left\lceil\min\left\{L_k,
     \max\left(1,\left\lceil\frac{B_{kp}}{D\sigma}\right\rceil\right)
     \right\}\right\rceil_2.
\end{aligned}
\label{eq:peer-demand}
\end{equation}
For $2\le|\mathcal P_k|\le a_k$, the local objective is
\begin{equation}
\begin{aligned}
 \underset{\{x_{kp}\}}{\operatorname{minimize}}\quad
     & \max_{p\in\mathcal P_k} B_{kp}/x_{kp} \\
 \text{subject to}\quad
     & \sum_{p\in\mathcal P_k}x_{kp}\le a_k, \\
     & x_{kp}\in\{1,2,4,\ldots,h_{kp}\}.
\end{aligned}
\label{eq:peer-surrogate}
\end{equation}
Algorithm~\ref{alg:gin-allocation} approximates Eq.~\ref{eq:peer-surrogate}
greedily. It gives each peer one channel, then doubles the count of the
eligible peer with the largest $B_{kp}/x_{kp}$. Each doubling must fit the
remaining budget and peer ceiling. Allocation stops when none fits,
preserving power-of-two counts of at least one per peer.

\begin{algorithm}[t]
\caption{Demand-balanced GIN channel allocation}
\label{alg:gin-allocation}
\small
\begin{algorithmic}[1]
\REQUIRE Active GIN peers $\mathcal P_k$, bytes $B_{kp}$, channels $K$,
connections $J_k$, FIFO settings $(D,\sigma)$, factor $f$
\ENSURE Local peer-channel ceilings $x_{kp}$
\STATE $a_k\gets\min(K,fJ_k)$
\IF{$\mathcal P_k=\varnothing$}
  \STATE \textbf{return} $\varnothing$
\ENDIF
\IF{$|\mathcal P_k|=1$}
  \STATE \textbf{return} $x_{kp}=\min\{K,\lceil a_k\rceil_2\}$
\ENDIF
\FORALL{$p\in\mathcal P_k$}
  \STATE Compute $h_{kp}$ using Eq.~\ref{eq:peer-demand}; $x_{kp}\gets1$
\ENDFOR
\STATE $s\gets|\mathcal P_k|$
\WHILE{$s<a_k$}
  \STATE $\mathcal E\gets\{p\in\mathcal P_k:x_{kp}<h_{kp},\ s+x_{kp}\le a_k\}$
  \IF{$\mathcal E=\varnothing$}
    \STATE \textbf{break}
  \ENDIF
  \STATE $p^*\gets\arg\max_{p\in\mathcal E}B_{kp}/x_{kp}$
  \STATE Break ties by the smaller peer rank.
  \STATE $s\gets s+x_{kp^*}$; $x_{kp^*}\gets2x_{kp^*}$
\ENDWHILE
\STATE \textbf{return} $\{x_{kp}:p\in\mathcal P_k\}$
\end{algorithmic}
\end{algorithm}

Endpoint negotiation makes each pair's ceiling compatible with both ranks'
allocations: $\overline c_{kp}=\min(x_{kp},x_{pk})$. The scheduler then chooses
payload-dependent counts within these ceilings. With a single peer, there
is no competing demand, so the allocator assigns the rounded-up budget,
capped by $K$. If peers outnumber the nominal budget, each retains one
channel and subsequent scheduling shares channels through queues.
Relay streams require consistent channels across their participants; \sys derives
their ceiling from the rounded-up GIN context count, with a fallback when
GIN is absent (\S\ref{sec:implementation}). The peer allocator and relay rule
thus supply the ceilings used to partition route streams.

\section{Experimental Methodology}
\label{sec:reproducibility}

\subsection{Motivation Workload}
Figure~\ref{fig:sync-model-size} uses veRL's fully asynchronous GRPO on
GSM8K, with separate 16-GPU training and rollout pools. Training uses
TP2/CP2/PP1, with EP8 for Qwen3-30B-A3B and EP1 for the dense models;
rollout uses four TP4 instances. Each update consumes 32 prompts with four
responses each. The plotted interval is trainer-side parameter publication,
including its internal barrier, divided in panel (b) by the complete
trainer update-cycle wall time, including waiting and computation.
Each point is one recurrent update after initialization.

\subsection{Communication Memory Measurement}
\label{sec:memory-measurement}

Figure~\ref{fig:memory-comparison} uses an eight-GPU TP4/CP2/EP8 trainer
on one host and two independent TP4 rollout instances on the other.
We subtract the loaded-model
baseline and validation copies from each rollout GPU's memory usage.
PyTorch live is allocated memory; cache is reserved minus allocated;
non-PyTorch is device-used minus PyTorch reserved. We take each GPU's
median resident increment across updates, then report the largest across
GPUs. All three components use the same subtraction and come from that GPU.

\subsection{Elastic Rollout Measurement}

In Figure~\ref{fig:elastic}, the membership timeline uses a single driver
clock and counts prepared rollout instances. Steady latency statistics
exclude the first update of each membership epoch.

\subsection{End-to-End RL Measurement}
\label{sec:rl-experiment-details}

The experiment in \S\ref{sec:eval-e2e} uses a TP2/CP2/EP8 Megatron
trainer and vLLM rollout. Configuration R1 in Table~\ref{tab:rl-e2e}
uses TP4/DP1/EP4 in four four-GPU groups; R2, R3, and R4 use
TP4/DP2/EP8, TP2/DP4/EP8, and TP1/DP8/EP8, respectively, in two
eight-GPU groups each. Runs contain 16 optimizer updates, each consuming
64 prompt groups with 16 completions per group.
Prompt and response lengths are capped at 2,048 and 8,192 tokens,
respectively, and the staleness threshold is 0.5. Paired response-token
counts differ by less than 0.5\%.

Throughput divides consumed response tokens by complete training-loop
time, including generation waits and final drain but excluding
initialization. Weight-update latency spans the full workflow from the
rollout-pause request through generation-resume acknowledgments.
Blocked rollout GPU time spans all-worker pause acknowledgment to the
resume request, multiplied by 16 GPUs; update work may execute during
this interval.

\subsection{Correctness Validation}
\label{sec:correctness-validation}

Our standalone harness retains reference copies of framework-loaded
parameters and clears destination tensors before each update. After
transfer, it checks every BF16 parameter for exact equality with its
reference, accounting for a changing final-norm marker that identifies
the expected version, and verifies that parameter addresses remain
unchanged. For FP8, an independent checkpoint-derived reference validates
block scales and quantized bytes, as well as parameters retained in BF16.
After restoring the original checkpoint values, the harness also compares
deterministic generation before and after the update sequence. Reference
construction, clearing, parameter checks, and generation are outside the
timed update interval.

\end{document}